# On the Lorentzian route to relativity

Atilla Gürel   e-mail: atillag57@gmail.com

and

Zeynep Gürel   e-mail: zgurel@marmara.edu.tr

**Abstract:** We discuss pedagogical problems associated with the conventional approach in teaching relativity and the potential value of "Lorentzian route to relativity" for addressing these pedagogical problems. We think that Lorentzian approach may be valuable part of an introduction to relativity but the historical route should be followed up to its end. It must namely include Einstein's application of Occam's razor in reinterpreting Lorentz transformations kinematically and the discussion of the central role of the evolution from Lorentz's "local time" to relativity of simultaneity. We discuss also what Occam's razor is not and why it should not cut "too deeply or too permanently".  Some misleading remarks in relativity text books concerning for example Kennedy Thorndike experiment and Trouton-Noble experiment indicate the not uncommon lack of awareness about the degree observational equivalence between prerelativistic (Lorentzian) and relativistic  viewpoints as far as special relativistic experiments are considered.

## 1. Introduction

There exist many studies aimed at particular aspects of special theory of relativity indicating difficulties in students understanding at all levels of education. A list of these papers can be found in the paper written by Arriassecq and Greca in 2010 where the authors propose historical and philosophical contextualization in an introductory course to STR. Although the teaching sequence they propose is designed for secondary education, the content of each item can be expanded/modified accordingly to suit the needs of higher education. A more recent study on the problems of preservice teachers with key concepts in relativity was conducted by Selçuk in 2011. The practical problem regarding a historical contextualization in science education is to find the right balance concerning the question to what extend history of science should be covered in science education. On one hand it is obvious that it is neither practically possible nor useful to get through all failed or outdated theories and explain why they have failed or are not popular anymore. On the other hand however, teaching well-established, modern theories straightforward by omitting their historical context may lead to problems of mental digestion, particularly if these new theories are associated with dramatic conceptual challenges defying daily common sense. If one considers that the associated conceptual changes have been not easily accepted by many scientists of their time and that there was a lot of discussion in scientific community before the new paradigm could become established, it is unfair to expect from the student to accept the new conceptual framework without going through a similar process of digestion. Vilanni and Aruda emphasized the parallelism between the two processes namely the nature of resistance to the new theory until its general acceptance and the student's difficulties in the learning process (1998). Thus, finding the optimum balance regarding the use of historical material is not a trivial task. However, the conventional way to introduce relativity is certainly highly problematic as we will discuss in section 2.

As known, Lorentz and contemporaries, specially Poincare anticipated that if there was a "conspiracy" of nature to prevent the discovery of the assumed preferred frame by experiments, than this "conspiracy" should be complete in the sense that all fundamental physical laws and not only laws of electromagnetism should have a Lorentz invariant form so that all fundamental laws "appear to be observationally valid" in all inertial reference frames. Thus, Poincare already anticipated a form of relativity principle (Hoffman, 1983, pp. 86) we may call "observational relativity principle" contrary to what we may call the "ontological relativity principle" where the phrase "appear to be observationally valid" is replaced by "are valid" as it was assumed for Newtonian mechanics under Galilean transformations and as it was restored for electrodynamics in particular and for relativistic physics in general by Einstein through a kinematic interpretation of the Lorentz transformations.

Among the relativistic effects of time dilation, length contraction and relativity of simultaneity, in our opinion relativity of simultaneity has a distinguished position within the historical evolution of the concepts. As known, concepts of time dilation and length contraction were notions that existed already prior to Einstein's 1905 paper although with a different meaning namely as dynamical effects of physical contraction of objects moving relative to the assumed preferred reference frame (Lorentz-Fitzgerald contraction) and slowing down of the motion of electric charges in reference frames moving relative to the preferred reference frame (Larmor dilation). It is the evolution from Lorentz's "local time" that represented reading of clocks located at different space points in the "*moving*[1]" frame (and that were assumed to be asynchronous in absolute sense), to Einstein's idea about relativity of simultaneity that restored the lost symmetry between reference frames and that made the kinematical interpretation of the Lorentz transformations possible. Thus the relativity of simultaneity being the true revolutionary conceptual step taken by Einstein in his 1905 paper, it would not be an exaggeration if we would call it the "essence of relativity". We think that a teaching strategy with historical contextualization has to take this fact into the account. Scherr et al. focused on students difficulties regarding the concept of relativity of simultaneity (2001). They proposed several activities to address the misconceptions related to relativity of simultaneity (2002). However the pedagogical problems we discuss in section 2 remain unsolved unless the justification for the conceptual change from pre-relativistic interpretation of Lorentz transformations to their kinematical interpretation is discussed in detail. To justify Einstein's conceptual step, we think it is necessary to demonstrate to the student that two clock synchronization processes namely "synchronization by light signal" and "slow separation after synchronization" would lead to the same result. In other words, it should be shown that if a clock is delayed by factor γ due to its velocity relative to the assumed preferred frame then clocks separated after synchronization in the *"moving*[1]*"* frame would have the same reading difference that they would have if one tries to synchronize the distant clocks in the *"moving"* frame by light signals so that there is no experimental possibility to verify their assumed asynchrony by checking the results of both methods against each other. Only after this point, the student can realize that unlike the simultaneity of co-local events the concept of distant simultaneity is not some information we can extract from nature directly and only then we can expect from the student to accept that we need a convention for distant simultaneity. Only then, we can expect from the student to accept Einstein's convention for distant simultaneity (which is in essence nothing but the postulate about one way velocity of light) as a convention best accordance with Occam's razor under given conditions.

---

[1] We will use the term "*moving*" (in italic in quotes) in the sense "moving relative to assumed preferred frame" throughout the paper.

In section 2 we discuss problems associated with conventional introduction to relativity. In section 4 we present a teaching strategy to address these problems. The use of Lorentzian route to relativity (that is called dynamical/constructive[2] approach in contrast to kinematical/principal approach of Einstein) for didactical purposes was already proposed by Bell. However, his treatment does not address the delicate issue of relativity of simultaneity that we consider as the "essence of relativity". In section 3, we want to make some remarks concerning a relatively recent debate on the subject. In section 5, we discuss some misleading remarks in relativity text-books regarding the experimental possibility to distinguish between the two viewpoints in context with the Kennedy-Thorndike and of Trouton-Noble experiments. In section 6, we discuss of what Occam's razor is not or why it should not cut too deeply or too permanently.

**2. Problems of Conventional Approach in Relativity Education.**

Chapters on relativity in most of the university physics text-books share more or less the following pattern:

1. The discussion electromagnetic waves, light as electromagnetic wave and the concept of ether that emerged in 19[th] century

2. Discussion of Michelson Morley experiment.

3. Lorentz Fitzgerald contraction hypothesis as a "last attempt to save the ether"

4. Derivation of Lorentz transformations starting with Einstein's postulates including the second postulate about the independence of light velocity from chosen reference frame.

After the derivation of Lorentz Transformations, the length contraction, time dilation, relativity of simultaneity are discussed. Sometimes relativity of simultaneity is discussed directly as consequence of invariance of light speed (the thought experiment with a train and two observers one insight and one outside the train )

There are following conceptual difficulties associated with this route.

   1. There is a subtle conceptual problem associated with step 4 that is in our opinion of high relevance from constructivist perspective of the learning process and that lies deeper then the obvious difficulty of accepting that the velocity of "something" can be the same in all reference frames: Until a student encounters relativity, time and space are fundamental concepts. They form the stage where things happen. Although it is clear that there is a need for "players" namely physical entities and events to measure space and time, time and space are perceived as entities that are there even if there is nothing that moves with a certain velocity. However, in the usual derivation of Lorentz transformations, the peculiarities (time dilation and relativity of simultaneity) of an entity (time) that is perceived as fundamental, is mathematically derived and thus tried to be understood as a consequence of some apparently peculiar behavior of an entity (light velocity) that is itself perceived not as fundamental because it is the attribute of one of the "players". Therefore when we derive Lorentz transformations starting with Einstein's postulates, because of the new distinguished position of light velocity as a new fundamental concept in a sense redefining *"how time and space behaves"*, the conceptual hierarchy that exists in students mind is damaged without actually being replaced by the new one. This approach tries to construct the new conceptual hierarchy without actually first fully deconstructing the fundaments of the former one namely Galilean space and absolute time. Therefore the student is left with confused conceptual hierarchy relationship, of which most probably he/she is not aware himself/herself. This reflects itself in the educational studies according to which both frameworks (Galilean space and time versus relativistic space-time) seem to coexist in students mind.

---

[2] The term constructive(referring to a theory or to an explanation) is used here in the sense of understanding macroscopic physical behavior as direct consequence of microscopic properties in contrast to a principal theory where the theory is build upon general principles even without the knowledge of microscopic details and imposes restricting conditions on possible constructive theories. It should not be confused with constructivist perspective of the learning process.

> They often attribute the relativity of simultaneity to the difference in signal travel time for different observers. In this way, they reconcile statements of the relativity with a belief in absolute simultaneity and fail to confront startling ideas of special relativity. (Sherr et al., 2001)

2. The name of Lorentz is mentioned in two different contexts. The first one is the contraction hypothesis; The second one is as the name of transformation formula that are derived in the books using Einstein's postulates. Thus it is natural that a student with inquiring mind may ask the following question :

*"That the relativistic coordinate transformations are named after Lorentz indicates that they must have already derived by him before Einstein. However the only mentioned major contribution of Lorentz mentioned in the book is "the proposal of length contraction for objects moving relative to ether to explain the null result Michelson Morley experiment without sacrificing ether concept". According to this hypothesis, although the total time for the light for two way path namely the apparent/measured average two way velocity is independent from direction for a chosen reference frame, the one way velocity is assumed to be c-v in direction of motion and c+v in the opposite direction in the frame moving through ether. Our derivation of LT however starts with the assumption that one way velocity is the same in both reference frames in all directions. How could Lorentz have derived these transformation formula that are only consistent with independence of one way light velocity from chosen reference frame and propagation direction, although he assumed in the contraction hypothesis that one way velocity depends on chosen reference frame in accordance with ordinary Galilean laws for the addition of velocities."*

He/she may ask further:

*"Beside this, there is a concept of time dilation and not only length contraction in Lorentz transformations, thus if Lorentz had derived these transformation formula before Einstein using some other assumptions that are different from Einstein's postulates, then certainly the contraction hypothesis alone couldn't have been sufficient as a starting point."*

3. The student may ask himself/herself "Why was the Lorentz-Fitzgerald contraction hypothesis discarded so easily if it could explain the null result of Michelson Morley experiment? Why should we accept something so strange like Einstein's postulate on invariance of light velocity that defies all common sense if there is a solution that does not contradict the common sense? There are statements like *"Einstein's solution to the problem was simpler or more elegant"* etc. But at this stage the student is not provided with enough information to appreciate the full value of the conceptual consequences of Einstein's revolutionary reinterpretation of the transformation formula, the mathematical form of which were already derived previously by Lorentz and Larmor independently. To convince the student of the necessity to jump from step 3 to step 4, some books dedicated to relativity mention other experiments like Kennedy Thorndike experiment (Resnick & Halliday 1992, pp. 21) (Sartori 1996, pp. 44-45) or Trouton-Noble experiment (Sartori 1996, pp. 46), stating that they cannot be explained from pre-relativistic point of view even if one includes the Lorentz Fitzgerald contraction hypothesis. Although this statement is true, it is highly misleading because Larmor dilation was as essential part of Lorentzian or pre-relativistic viewpoint as the length contraction hypothesis was. We will discuss these experiments in section 5.

### 3. Constructive[2]/Dynamic Approach versus Principle/Kinematic Approach to Relativity.

*"Does the length of a moving rod contract as a consequence of equations that describe the dynamics of its microscopic constituents, or does it contract because of Minkowski geometry of space –time?"*

Simply put one may say, accepting the first answer for the above question means defending dynamical/constructive approach to relativity while accepting the second answer means advocating for kinematic or principle approach. However, as we will see, this is a too imprecise differentiation of the two viewpoints.

In 1976 Bell has proposed to adopt the Lorentzian route to Lorentz transformations as an alternative approach in education where the length contraction and time dilation are presented as dynamical effects both being a consequence of Maxwell equations (Bell, 1987). In 2009, Wiener used quantum mechanics to develop a more realistic model to achieve the understanding of microscopic foundations of length contraction as a consequence of dynamics.

Brown and Pooley defend the constructuve/dynamical approach to relativity and unlike Bell who considered this approach for didactical purposes, they consider it as a challenge for the established understanding of relativity (Brown & Pooley 2001, 2003; Brown 2005). This started a debate (Balashov & Janssen 2003; Janssen 1995; Norton 2007; Janssen 2009; Nerlich 2008) on the subject.

Let's explain first why the question at the beginning of this section in quotes is not precise enough for a proper differentiation of the two viewpoints:

While defending the kinematical approach against a dynamical approach, Nerlich writes for example:

> A moving rod, however it is constituted, is contracted in the direction of motion by known factor; a moving tensor field, however complex, is equally contracted (i.e. the tensor components of each electron field change) by the same factor in direction of motion. The one contraction <u>does not cause</u> (underlining by us) the other (Nerlich, 2010).

However from relativistic point of view for an observer in any inertial frame, there is nothing wrong with the following statement:

*"Maxwell equations are valid in my reference frame. The electromagnetic field of a moving electric charge (moving relative to the inertial frame of the observer) has this particular form (a contracted electric field together with a circular magnetic field)  <u>because</u> it is the correct solution to Maxwell equations for a uniformly moving charge"*

Thus, it is not wrong to conclude for the same observer that the moving (moving relative to the observer) rod contracts, *because* atomic orbitals are squeezed, *because* of these changes in the field of the atomic nuclei. When we take a look at the concept-map (App.1), the relativistic connection from 3 to 19 represent the kinematical explanation which sounds trivial. However, this doesn't make the long route to 19 over 13 and 14 (which was the only possible route to 19 in the pre-relativistic view) totally obsolete. The critical connection that is specific to relativistic view is the connection from 2 and 3 to 6 and 9. Once this connection is established, the long route to 19 over 13 and 14 and the short route from 3 to 19 are not two competing routes, one excluding the other anymore, but they are two consistent faces of one and the same reality according to relativistic picture. This is also the opinion of Dieks (1984) and Wiener (2009). There are two aspects in which the relativistic view regarding the long route (dynamical explanation of length contraction and time dilation) differs from Lorentzian view: The long route is a valid description in any inertial frame and not only in some assumed preferred frame(difference between 11 and 12) . It is valid not only for the length of objects, the microscopic parts of which are hold together by electromagnetic forces but it is valid for objects of any type (the solar system as a whole or even the size of individual elementary particles that seem to have an internal structure) and it is valid not only for duration of electromagnetic processes but duration of processes of any type. In fact it is only relativistic view that guarantees that if we have discovered correct dynamical equations and if we follow the long route and if we make our correct calculations we must be able to achieve our destination 19 over the long route.

This is an important prediction by relativity because until advent of quantum mechanics we could anticipate but not satisfactorily complete the long route for the contraction of the length of a solid body since it was impossible to explain the finite size of the atoms because an electron orbiting the nucleus would radiate according to Maxwell equations. There may be still phenomena where the dynamical laws and related calculations for the long route may not be as well established as in the case of electronic states of the atom (like the dilation of half life of mesons for example prior to development of quantum chromo-dynamics). However it is our confidence in special relativity that allows us to believe that *if we find correct dynamical description of these phenomena and if we carry out correct calculations/simulations with them, we should be able to follow the long route up to its end.* We will discuss this in context with Trouton-Noble experiment in section 5.

Thus, since the kinematical approach does not exclude the validity of dynamical reasoning, in order to precisely differentiate the purely constructive approach from principal/kinematic approach the question at the beginning of the section must be slightly modified. The definition of constructive approach must contain an explicit rejection of the Minkowski geometry for space-time. Thus, one may be inclined to modify the question in the following way:

*"Does the length of a moving rod contract as a consequence of Lorentz invariant equations that describe the dynamics in a Galilean space and time, or does it contract because of Minkowski geometry of space-time?"*

The first alternative now represents the neo-Lorentzian view that assumes the existence of a hidden preferred frame. We may ask now whether there may be any other way to prefer the constructive approach without defending the neo-Lorentzian view?

If one reads Brown's following statement in the preface of his book "Physical relativity" where he defends constructive approach,

> I must emphasize from the outset that this approach does not involve postulating the existence of a hidden preferred inertial frame! The approach is not a version of what is sometimes called in the literature the neo-Lorentzian interpretation of special relativity. (Brown, 2005) one wonders how this may be possible. One finds the answer in the section 2 of (Brown & Pooley, 2006). They defend a relationist view of dynamics based on Machian framework where space is not an entity by itself acting as a stage for dynamics but where it represents merely reflection of relation between bodies/fields (Pooley.O & Brown H.R, 2002). Thus it is not only "the Minkowski-spacetime" that they consider "as a glorious non-entity" as the title of their paper (2006) lets suggest, but it is "space and time" of whatever geometry. On the other hand however in the preface of "Physical Relativity" on page ix Brown writes :

> I should perhaps clarify that my book is not designed to be a defense of a Leibnizian/Machian relational view of space-time of the kind Barbour has been articulating and defending with such brilliance in recent years, and in particular in his 1999 The End of Time. Although I have sympathies with this view, in my opinion the dynamical version of relativity theory is a separate issue and can be justified on much wider grounds, having essentially to do with good conceptual house-keeping. (Brown, 2005)

However, there are only two ways for defending the purely constructive/dynamical version of relativity: Either the neo-Lorentzian approach within Galilean space and time or strict Machian approach rejecting the conventional notions of space and time completely. There is no third way in between. The discussion of substantivalist and relationalist views of space and time is beyond the scope of this paper. We will discuss the neo-Lorentzian approach in sections 5 and 6.

**4. How to Teach Relativity.**

Since Michelson Morley experiment establishes only the invariance of the "observed average two-way velocity of light", jumping directly from this point to Einstein's postulate about the invariance of one way velocity of light is nothing but demanding from the student to accept something by heart without sufficient justification. Referring to experiments like Kennedy-Thorndike experiment and Trouton-Noble experiment as evidences falsifying Lorentzian viewpoint and justifying relativistic viewpoint (thus Justifying Einstein's postulate about one-way light velocity) as it is done in some text books dedicated to relativity is highly misleading as we will discuss in section 5.

Nerlich says:

> Thus the Lorentzian pedagogy as Bell presents it, nowhere unequivocally draws on dynamics. Nor can it restore confidence in perfectly sound and useful concepts already acquired. (Nerlich, 2010)

The problem here is whether we are sure that the students have really acquired them. Although most of the students can use the relevant formula and solve the problems without difficulty, the educational research, mentioned in the introduction indicates it is questionable whether the students have "acquired the relativistic concepts" at all.

We think incorporating Lorentzian route to Lorentz transformations is highly valuable to address the didactical problems presented in section 2 and to develop an in depth understanding of relativity. It would allow to understand the mathematics of the transformation formula without first giving up familiar notions about space time and without the necessity to rely upon the counterintuitive postulate about light velocity. However the Lorentzian pedagogy as it is formulated in Bell's paper has a missing element from an educational point of view: When he says that they (the Lorentzian route and Einstein's route to special relativity) differ only in philosophy, it sounds as if it is merely a matter of taste which of them, one may prefer. We think a historical route would be incomplete without the presentation and justification of Einstein's reinterpretation of Lorentz transformations and discussion of its advantages regarding further predictions like for example dilation of the period of gravitational pendulum or half life of mesons etc. that could not (and still cannot) by a theory based upon Lorentzian view.

While discussing origins of STR, Arriassecq and Greca write :

> There was a need to have a new viewpoint to solve the theoretical contradiction (underlining by us) that arose when Maxwellian electrodynamics gave an explanation for the effects of motion between magnet and a conductor, depending on which of the two is at rest or in motion. That need was the starting point of the development of the SRT. (Arriassecq and Greca, 2010)

Similarly, Nerlich writes:

> It was not the general assumption of the principle of the relativity for laws of nature but the problems of Maxwell electrodynamics (underlining by us) as Einstein tried to clarify in the discussion of the relative motion of a magnet and a conducting coil and the different descriptions of the same physically observable effect(the magnitude of the induced current in the coil) depending on the chosen reference frame that was the starting point for Einstein. (Nerlich, 2010)

However it is not appropriate to call it a "theoretical contradiction" or consider it as a real "problem" of a theory like being in contradiction with an experiment but it is rather an esthetical unease about an asymmetric explanation of the phenomena despite the fact that the resulting current depends only on relative velocity of the objects. Immediately after discussion of magnet and coil, Einstein writes namely:

> Examples of this sort, together with the unsuccessful attempts to discover any motion of the earth relatively to the "light medium," suggest that the phenomena of electrodynamics as well as of mechanics possess no properties corresponding to the idea of absolute rest. (Einstein, 1905).

Thus, for Einstein the discussed experiment and the null result of experiments designed to detect the ether wind belong to the same class of facts namely facts supporting the argument that it is only relative motion that is physically relevant and that there is no evidence supporting the concept of absolute motion. Just before 1905, the concept of mechanistic ether was not considered as absolute necessity as it was considered in 1880's. Poincare already had expressed his opinion according to which the ether might be useless[3] (Granek, 2001) (Poincare, 1902, pp. 180, 215). Electromagnetic waves propagate in all directions with same speed only in a reference frame where the Maxwell equations are valid. Thus as in the case of coil and magnet, the investigation of the possible effects of the motion of earth on observed velocity of light need not necessarily be interpreted as attempts for empirical verification of a hypothetical material medium but the question here is whether Maxwell equations are valid only in a single preferred frame (as it should be if Galilean transformations are valid) or whether they are valid in all inertial frames, namely the question is ultimately about the principle of relativity. Thus the necessity of a preferred frame does not emerge necessarily from the need for a material medium to understand propagation of electromagnetic waves but it emerges from the incompatibility of Maxwell equations with Galilean transformations. Thus we don't think that there is much didactical benefit in separating the role of discussion of magnet and coil from the role of negative results of ether detection experiments.

Arriassecq and Greca state:

> Suggesting that Einstein used Michelson's experiment as starting point for developing the SRT, helps to create in students a distorted view of scientific activity, favoring in this way a completely empiricist conception of science. (Arriassecq and Greca, 2010)

They mention also that there is evidence confirming that the role of Michelson Morley experiment played in the genesis of SRT has been minor and indirect. Therefore we don't know to which experiments exactly Einstein refers by the phrase "*unsuccessful attempts to discover any motion of the earth relatively to the light medium*". However considering that Michelson Morley experiment is the most well known experiment of this type, we think that it is not improper to give it a central role in an introductory discussion. After all, the didactical route cannot contain all the elements in the historical route, rather some carefully chosen elements in the historical route should be put such a way together so that maximal didactical benefit can be obtained. We think that keeping the central didactical role of Michelson Morley experiment do not *create a distorted view of scientific activity, favoring in this way a completely empiricist conception of science*, if it is put into the right context in the teaching sequence as described in the previous paragraph.

We must mention also a little misleading interpretation of the pre-relativistic view of the coil and magnet when comparing the two cases where either they are both at rest or they are both moving with the same velocity through the assumed ether.

> Since the charges of the moving conductor are themselves moved through the magnetic field, that field also exerts a force on them and produces a current. The two currents, one due to the induced electric field, the other due to the motion of the charges in the magnetic field are in opposite directions and turn out to cancel exactly. In both cases, there is no measurable current. (Norton J.D. www.pitt.edu)

There are no two currents canceling each other in the pre-relativistic view either but there are two forces of same magnitude acting in opposite directions so that the net *resulting force* acting upon a charge is zero so that the charge cannot move at all, so that there is no current at all in pre-relativistic view either. Since the concept of "current" is associated with "motion" of charges, and since net zero current can be produced by many ways (for example when same amount of charges moving in opposite directions or when charges of opposite sign moving in the same direction) and since a physical motion can lead to other physical effects like heat production by the Ohmic resistance etc., the above interpretation "with two opposite currents" is a little misleading depiction of the pre-relativistic view.

The lesson should consist of two main parts. In the first part the term preferred frame should be used to refer to the frame where Maxwell equations are valid. Any reference to the concept of ether as a material medium and discussions about its "contradicting" properties may be omitted at this stage because of the arguments explained in the paragraph above immediately following our quotation from Einstein's 1905 paper in this section and in order to use time more economically. Following simple argument is sufficient to convince the student that Maxwell equations cannot be Galilee invariant: Since there is only one propagation velocity associated with the wave solutions of Maxwell equations in vacuum, it is obvious that if Maxwell equations are valid in a particular frame, then according to Galilean velocity addition rules the propagation velocity cannot have the same value in all directions in frames other then this particular frame which means that Maxwell equations cannot be valid in these reference frames. Thus if Galilean transformations are correct then there can be only one preferred frame where Maxwell equations can be valid.

Speaking in the language of the concept-map (App.1), the first part of the lesson should start with 1,4, and 5 and then should follow the route over 13 and 14 to reach the destination 19. In the second part arguments for Einstein's reinterpretation of Lorentz transformations should be introduced and the advantages of Einstein's approach should be discussed.

---

[3] Thus we prefer to use the term "pre-relativistic" instead of the term "etheristic".

Following points should be the key considerations in the first part:

- Discussion of Michelson Morley experiment and its null result as usual followed by the discussion of the Lorentz Fitzgerald hypothesis. However it should not be presented as a hypothesis brought from the air without any justification just to explain the null result as it is done in some text books. Instead the "deformation" of the electromagnetic field of a point charge (for example the nucleus of an atom) as a consequence of a uniform motion according to the Maxwell equations should be mentioned so that a physical contraction of atoms and consequently contraction of a solid object appears as a reasonable idea.
- Michelson Morley experiment proves that the observed (apparent) average two way velocity namely the total travel time for the light signal is independent of the chosen reference frame and propagation direction. However, it doesn't allow us to make any decisive statement about one way velocity of light. We don't intend to give up right at this point and we want to try to determine the one way velocity of light in opposite directions.

  We need distant synchronous clocks to measure one way velocity of light. Since the velocity of the light signal cannot be used as a synchronization method under Lorentzian viewpoint, the only remaining method is the separation after synchronization. We need to show that unlike the clocks that are separated after synchronization in the preferred frame, clocks that are separated in a moving frame would display different times if there is a clock dilation that depends on the velocity of the clock relative to the preferred frame even in the limit case when the separation velocity goes to zero. As Miller states (2010) for an exact derivation of the formula for Larmor dilation to show that motion of electric charges in a moving frame are slowed down by a factor $\gamma$, one needs relativistic mechanics. However we need first Lorentz transformations to justify the use of relativistic mechanics so that we seem to get into a vicious circle. However, Millers way of addressing this difficulty is in our opinion problematic from didactical point of view[4]. We can use a pedagogical trick at this point to circumvent this difficulty. Since it is obvious from discussion of Michelson Morley experiment that a *"moving"* light clock would delay by factor $\gamma$ independent of the orientation of the mirrors, we can use two light clocks to see what happens if we separate them after synchronization in the moving frame. We may postpone the discussion of behavior of different types of clocks to the second part of the lesson. A simple calculation shows (App. 2) that there is an inevitable clock reading difference between the two clocks that have been separated after synchronization in the *"moving"* frame because of the difference between their absolute velocities during the separation and that this resulting time difference in the limit case when separation velocity goes to zero has exactly the value that compensates travel time difference for light in the *"moving"* frame in opposite directions so that the one way velocity of light in opposite directions measured with these slowly separated distant clocks would appear to have the same value c. Thus it appears that the *"moving"* observer with light clocks is in a sense "desperate" situation because he/she has no possibility to verify the asynchrony of his/her distant clocks if there is an absolute clock dilation combined by an absolute length contraction. Thus, the student realizes here that for a *"moving"* observer it is not only impossible to verify the absolute motion by a Michelson Morley type experiment but it is impossible to verify this assumed absolute motion even if one tries to measure the one way velocity of light.

---

[4] Asserting equation 7 immediately after equation 6 in (Miller, 2010) is problematic because since formula 6 is derived from the dynamic description of length contraction, using the inverse formula 7 without further justification means automatically telling the student that the length of a rod in S is contracted when viewed from S'. However accepting this type of symmetry is as counterintuitive as the postulate about the invariance of light speed. Telling the student *"if we repeat similar argument from the point of reference frame S' "* is insufficient because it is unclear how we can be allowed to "repeat a similar argument" from the viewpoint of reference frame S' where the rods have shown to be definitively shorter than the rods in S. When we want to develop an alternative route to solve pedagogical problems mentioned in section 2, neither time dilation nor length contraction can be presented as symmetric phenomena right at the beginning because these are as counterintuitive statements as Einstein's postulate on light velocity. The symmetry can be restored only when relativity of simultaneity is introduced and discussed as replacing the Lorentzian concept of "local time" in *"moving"* frame.

This presentation has following didactical benefits:

- The student realizes that contrary to the simultaneity of co-local events, the information about whether two distant events are simultaneous or not, can not be extracted from the nature directly by observation without some prior assumption. This insight is an important step to mentally prepare the student to the counterintuitive concept of relativity of simultaneity and to the meaning of the reinterpretation of the transformation formula by Einstein.
- The mathematical basis to understand the difficult concept of relativity of simultaneity namely that clock reading at different positions may have different values is not derived as a consequence of Einstein's postulate about light velocity which is difficult to accept just on the basis of the results of Michelson Morley experiment but one peculiarity of time (its dependence on location in the "moving frame") is derived as the consequence of another peculiarity of time(dilation) to address problem 1 in section 2.
- It stimulates the student to ask whether there might be other possibilities to detect the absolute motion or whether for example all clocks would be delayed at the same rate as light clocks etc. Thus it prepares the student for the discussion on the difference regarding the degree of generality of the two alternative viewpoints.

- Lorentz transformation formula for $x' = \gamma(x - vt)$ can be understood as the consequence of Lorentz-Fitzgerald contraction. Lorentz transformation formula for time $t' = \gamma(t - xv/c^2)$ can be derived as the consequence of the time dilation (App. 2) combined by length contraction so that there is no need to use the counterintuitive concept of invariance of light speed at the very beginning that is unjustifiable on the result of Michelson Morley experiment alone.

Having completed the "derivation part", we can precede to the second part of the lesson namely the discussion part. This part should contain following topics.

- Derivation of the inverse of Lorentz transformation to show that inverse of Lorentz transformations are also Lorentz transformations which means that the light clocks in preferred frame must appear to be delayed when compared with moving clocks and rods in the preferred frame appear to be contracted when compared with moving rod if time is measured by the asynchronous light clocks in the moving frame. Thus, despite the assumed physical asymmetry between the considered reference frames, this asymmetry remains hidden if there is no way to verify the asynchrony of the moving clocks. We should mention here that *the transformation formula themselves only relate coordinates and time of one frame with the coordinates and time of the other frame and that they do not contain any information about in which frame distant clocks are synchronous and in which frame they are asynchronous. In other words the Lorentz transformations do not contain any restriction regarding the position dependence of clock readings within a frame namely regarding the mathematical form the functions t(x) or t'(x'). We are allowed to choose one of these two functions freely and then to calculate the form of the other one. If we choose t(x) = t, namely if we assume that t is independent of x then t' is a linear function of x' and vice versa (eq. 4 in App. 2).*

- We can return here to the question about whether all possible type of clocks would delay at the same rate as the light clocks. We think that in an introductory course there is no need for an exact relativistic derivation of Larmor dilation. It is enough to mention that calculations showed that the motion of electric charges under electromagnetic forces in a *"moving"* frame is slowed down by the factor γ because of the change of the electromagnetic field of a uniformly moving charge. Then we can discuss the possible consequences of Larmor dilation for material clocks where the working of the clock is ultimately based on electromagnetic processes as in a mechanical clock with spring or in a quarz clock. At this point, it should be discussed what type of clocks are not of electromagnetic origin (pendulum, sun-clock). The students are encouraged to think about the possible consequences of the following two alternatives:

    1. *What if moving gravitational clocks (or generally clocks that work not through electromagnetic forces) are not affected by their motion relative to the assumed preferred frame?* (This would

mean that there is a potential possibility to detect the preferred frame just by comparing electromagnetic clocks and gravitational clocks).

2. *What if they are delayed at the same rate as the electromagnetic clocks?* (The principle impossibility of detecting the assumed preferred frame by comparison of clocks and rods or light velocity measurements).

Here we could mention that in the pre-relativistic period, people anticipated the possibility that all type of clocks would be delayed at the same rate and that they tried to understand this as a consequence of the possibility that all types of forces would ultimately be the manifestations of electromagnetic forces. They tried to calculate the impact of Maxwell equations on the motion of electric charges and tried to modify laws of the mechanics (of charged particles) accordingly.

- Einstein was not comfortable with the view that the laws of nature are arranged so that they hide a physical truth from the observers. This is the context where he said "God is subtle but not malicious"[6] where he meant *"the laws of nature may appear to be elegant and well designed but this doesn't mean that they may have such a form that it seems as if they are cleverly designed just to hide a truth (the preferred frame) from us"*. It seems, for Einstein the symmetry between the reference frames regarding the coordinate transformation formula and the principle of relativity were too valuable to assume that they were merely an observational and illusory byproducts of dynamic equations. Symmetry reflects namely some type of order or even beauty. If searching for fundamental laws of nature is a search for some deeper order and simplicity in the complicated and chaotic world of phenomena, then how could one consider the empirically established symmetry as an illusion? He must have seen this as a setback after centuries of Galilean relativity. Einstein's 1905 paper contains a discussion of distant simultaneity at the very beginning just after a short discussion of experiments that seemed to make the concept of absolute motion superfluous. Thus, independent of the question how much Einstein actually was informed about the work of Larmor and Lorentz, the effective message of the paper to people preferring the Lorentzian approach is the following: *"What is the price we have to pay, if we want to reinterpret Lorentz transformations as reflecting real symmetry between two inertial reference frames instead of interpreting them as formula describing the distortion of rods and clocks because of their absolute motion. Namely what prevents us from declaring the clock readings in what you call "moving" frame to represent the actual time for an observer in this frame. Obviously it is the concept of the relativity of simultaneity that seems so difficult to accept. But is there any physical reason that forbids us to assume that simultaneity may be relative? It seems it is nothing else but a prejudice about time that we developed through daily experience."*

Thus the advantages of Einstein's viewpoint seemed to be the following:

1. The reason for observational validity of relativity principle is its real validity as a principle of nature (connections from 2 to 9 or from 3 to 19 in the concept-map (App.1)) and not being merely a "byproduct" of dynamic equations. Thus the conceptual structure is better in accordance with Occam's razor.

2. Einstein's view does not refer to an entity (preferred frame) that is experimentally principally undetectable according to the theory although it is assumed to be something real.

3. Since Einstein's viewpoint redefines the space time geometry and establishes the general ontological validity of relativity principle as a fundamental principle of nature, its consequences are not restricted to electromagnetic phenomena. It demands that *all* fundamental laws have to be Lorentz invariant. This principle has been proven itself very fruitful since then in many cases like for example in development of special-relativistic mechanics in a straightforward way by Einstein, or much later in developing the relativistic quantum mechanics and relativistic quantum electrodynamics by providing the Lorentz invariance as a precondition. The construction of space-time metric in accelerated frames derived from special relativity by considering acceleration as infinitesimal jumps between inertial reference frames and then applying these results to describe the effect of the gravitation on the metric by using the principle of equivalence led to development of general theory of relativity which has proven itself as the most successful theory to describe the behavior of matter at astronomic or cosmic scales.

---

[6] The English translation of the German word "raffiniert" as "subtle" is not precise enough in our opinion. The word "subtle" has meanings like deeply hidden, not easily noticeable etc. The term "raffiniert" when describing an act like finding a solution to a problem or designing something has the meaning "well thought, elegant". Here, in our opinion, the word "sophisticated" would reflect the original meaning better.

## 5. Some Misleading Remarks in Relativity Text Books

We want now address some not uncommon misconceptions as they reflect themselves in relativity text-books, regarding the degree of observational equivalence of neo-Lorentzian viewpoint and the relativistic viewpoint as far as special relativitistic phenomena are considered [5].

We mentioned in Section 2 that in some text-books, Kennedy Thorndike experiment (Resnick & Halliday, 1992, pp. 21) (Sartori, 1996, pp. 44-45) and Trouton Noble experiment (Sartori, 1996, pp. 46) are mentioned as examples that contradict the prerelativistic view. It is stated in (Resnick Halliday 1992, p.21) for example that although etheristic view including contraction hypothesis of Lorentz and Fitzgerald could explain Michelson Morley experiment it fails to explain Kennedy–Thorndike experiment. A statement that is although being true by itself, simply ignores the fact that the idea of Larmor dilation was proposed long before the Lorentz-Fitzgerald contraction hypothesis and that immediately before 1905 it was as essential part of pre-relativistic view as Lorentz Fitzgerald contraction. As the name of the paper indicates the Kennedy Thorndike experiment verified Larmor dilation (Kennedy and Thorndike, 1932). Ironically, despite the fact that it is presented in the listed books above as an experiment contradicting Lorentzian view, it is one of the rare sources where one can find the derivation of Lorentz transformations in their Lorentzian interpretation namely based on Lorentz-Fitzgerald contraction and Larmor dilation without the Einstein's postulate on light velocity. It was this paper that originally inspired us to write this article.

Now let's discuss the Trouton-Noble experiment that was proposed to detect a supposed motion relative to ether. A system of two opposite charges that are connected by a rigid insulator (typically a capacitor that consisting of two oppositely charged plates) are moving with velocity v in frame S. The angle between the direction pointing from one charge to the other and the direction of motion has some arbitrary value between 0 and 90. If we observe everything from the reference frame S' that is co-moving with the charges, from relativistic point of view there is no magnetic field because charges appear to be at rest in S'. Therefore there is only electric attraction between the charges and there can be no torque that tends to rotate the capacitor. Since, in the pre-relativistic view, magnetic fields are considered to be created by the motion of charges relative to preferred frame (and not relative to observer), it has been proposed that the lack of torque demonstrates that the pre-relativistic view can be considered as experimentally refuted. The conducted experiment revealed no torque thus confirms the relativistic view.

However the observed null result leads to the following question:

*"How do we explain the missing torque relativistically when we describe everything in S despite the existing magnetic field in S?"*

In other words how do we explain the missing torque by the long route in the concept-map (App.1).

However another question implicitly follows this one whether it is formulated explicitly or not:

*"If we can explain the missing torque from the viewpoint of an observer in frame S despite the presence of magnetic field in S, how can we claim that the missing torque disproves pre-relativistic view? Because the correct explanation would also explain automatically why there is no torque in pre-relativistic view"*

The solution is a simple straightforward application of relativistic mechanics. It is a remarkable phenomenon, that it took almost 100 years to find the correct simple solution to the problem. The solution of the problem is that, although there is a torque, namely although there is a "magnetic force **v** x **B** " so that the resulting force **dp**/dt is not aligned with the line that points from one charge to the other, this torque does not tend to rotate the system of charges because unlike in Newtonian mechanics, the direction of dp/dt is not the same as the direction of **dv**/dt in relativistic mechanics (Franklin 2006). Now see, "surprisingly", the calculation shows that the direction of **dv**/dt "accidentally!" coincides exactly with the direction of the line that points from one charge to the other so that there is no tendency to rotate. A conspiracy of nature? No! Just a natural consequence of Lorentz invariant equations. Thus, the null result of Trouton-Noble experiment doesn't falsify pre-relativistic viewpoint but it tells us merely that it is not only laws of electrodynamics that are Lorentz invariant but laws of mechanics should be Lorentz invariant, too. But, as we mentioned in the introduction this was not something that Lorentz or Poincare didn't already anticipate however the approach adopted in pre-relativistic period to solve this problem was trying to modify laws of mechanics (for charged particles subjected to electromagnetic forces) by taking the consequences of Maxwell's equation into the account.

The following case represents another striking example about the lack of awareness about the degree of observational equivalence of both viewpoints.

Wolfgang Rindler writes in his book "Essential Relativity"*:*

> For a while the Lorentz theory provided an alternative to Einstein's theory, equivalent to it observationally and less jolting to classical prejudices. But it was also infinitely less elegant, and above all, less suggestive of new results. Though apparently based on a preferred frame, the Lorentz theory yields symmetry, as far as observable predictions go, between all inertial frames. For from the assumed length contraction and time dilation relative to the ether frame it follows that rods and clocks moving at speed v through any inertial frame appear, respectively, to be shortened and to go slow relative to that frame by the same factor (Rindler, 1977, pp. 7)

These are statements we agree upon. It is surprising that after 25 years, Rindler writes in "Relativity , Special General and Cosmological" :

> Thus electromagnetic theory was left with a serious puzzle: The average to-and-fro light speed in a given ether wind is direction-independent.(Modern laser versions have confirmed this experiment to an accuracy of one part in $10^{15}$ ) The Michelson-Morley result is short of <u>what we know today, namely one-way speed of light</u> (underlining by us) in all times is independent of ether wind. This is nicely demonstrated by the workings of international atomic time, TAI (Temps Atomique international). TAI is determined by a large number of atomic clocks cluctered in various national laboratories around the globe. Their readings are continuously checked against each other by radio signals. (Rindler, 2001, pp. 10)

It is obvious that clocks synchronized according Einstein's convention for distant simultaneity (namely either by light signals or by slow separation after synchronization both leading to the same result) would measure the value c for one way velocity in any reference frame. Thus, contrary to above remarks we still don't have any directly established knowledge about one way velocity of light. We just adopted a convention of distant simultaneity proposed by Einstein in 1905 paper, that was best in accordance with Occam's razor, and that is consistent with the assumption that one way velocity of light is c in all frames in all directions. And this assumption has been proven itself as fruitful as we explained in the last paragraph of the previous section.

**6. What is Occam's razor not and why it should cut not too "deeply" or too "permanently"**

When we look at the three arguments at the end of the Section 4 that make Einstein's approach preferable, at first look, argument-2 (avoiding to refer to a principally experimentally unverifiable entity in the theory) may seem to be already contained within the argument-1 (the simplicity of logical structure), but this is not so. A theory may refer to a principally undetectable entity, thus it may be less preferable according to argument-2 but referring to this entity may make the structure of a theory simpler. The quantum mechanical wave function is a good example of such an entity. According to Copenhagen, interpretation the wave function represents a probability amplitude which allows us to calculate the probability of a particular outcome in a measurement. The wave function in its entirety can never be observed directly but its state can be concluded only indirectly after statistical evaluation of large number of measurements on identically prepared systems. Yet QM is a good theory because through it a wide range of phenomena that couldn't be understood at all, became understandable. The simpler conceptual and logical structure (Argument-1) is a more important issue than the question whether a theory refers to an undetectable entity or not (Argument-2).

Occam's razor as it is expressed in the form "for a given set of experimental facts the simpler explanation is preferable" is in truth a special case of a more general criteria that determines the quality of a theory that we may express in form of the following ratio:

6.1     DP/CA

Where DP is "Domain of the phenomena successfully described by the theory" and CA is "the complexity of the axiomatic structure of the theory" like the total number of fundamental independent concepts or entities in the axioms, the total number and the complexity of axioms.

The greater this ratio, the better is a theory. However since it is obvious that there is no way to exactly define and measure neither the nominator nor the denominator quantitatively it merely represents a qualitative argument. In some cases the previous theory is included in the new theory as a limit case as Newtonian theory is contained within the relativistic mechanics as the limit case when the velocities are so small when compared to light velocity so that relativistic effects can be neglected. In such a case it is obvious that the DP of the new theory is larger than the DP of the old theory and even if the axiomatic structure of the new theory is a little more complex than the axiomatic structure of the old theory the success of the new theory to explain phenomena that couldn't be understood in the old theory at all makes the new theory favorable. In this sense argument-3 together with argument-1 at the end of section 4 represent the real power of Einstein's view when compared to Lorentzian view. Einstein's view provides a clear and simple answer to the question why all fundamental laws and not only laws of electrodynamics should be relativistic (greater nominator "DP"). The answer is namely "because of the Minkowski geometry of space time which is much simpler when compared to incomplete results of the tedious efforts of dynamical unification, (thus smaller denominator "CA"). However things are more complicated when for two theories the two domains represent two intersecting sets. This is the case for example for general relativity and relativistic quantum mechanics. General relativity is better at astronomic and cosmic scales and quantum mechanics is better at microscopic scales. Both theories have classical special relativistic mechanics as a limit at ordinary macroscopic scales (for size of objects and distances of the order of magnitude of meters) since on one hand, the curvature of space can be neglected and on the other hand relativistic quantum mechanics becomes classical special relativistic mechanics by a relativistic equivalent of Ehrenfest theorem for expectation values of operators[7].

According to neo-Lorentzian view, the space time is Galilean but all fundamental laws of physics are Lorentz invariant because of some common origin of inference that is still waiting to be discovered. It is clear that any theory based upon the neo-Lorentzian view can be regarded as an alternative to relativistic viewpoint only if it could achieve the complete unification of dynamics. However would even such a unification make the Minkowsi space time superfluous as Einstein's reinterpretation of Lorentz transformations made once the concept of ether to appear superfluous ?

Jannsen's viewpoint on this subject is the following:

> We can imagine that Minkowski space-time will emerge in the low-energy limit of some future theory of quantum gravity that does not include any spatio-temporal notions among its basic concepts. Such a theory would provide an answer to the question 'Why Minkowski space-time?'. There is no reason to think, however, that this deeper theory would require us to move any of the phenomena in which Lorentz invariance manifests itself from the column of kinematics to the column of dynamics as established by special relativity. (Janssen, 2009)

---

[7]Things are actually a little more complicated for macroscopic limits of quantum mechanics since it must involve also the process of decoherence too and not only relativistic equivalent of Ehrenfest theorem to understand classical behavior as macroscopic limits of quantum mechanics and this issue is still not settled to everyone's satisfaction but this is beyond the subject of this paper.

Of course even a complete dynamical unification does not automatically mean that we can throw away the concept of Minkowski spacetime but such a development would make it appear more as a "matter of taste" then it appears now. Even in this case, if one considers the "ontological principle of relativity" as superior to "observational principle of relativity", a Minkowski space time would still be the choice. However, the answer would also depend on the specific nature of the deeper theory, namely whether the concept of a preferred frame is explicit and inseparable part of the new deeper theory. To clarify this lets consider the following hypothetical case:

Obviously we still don't have a successful (experimentally verified) theory that explains the masses of elementary particles consistent with all their quantum properties. Basically there are two ideas:

1. The idea of Higgs bosons as an add-on mechanism to known particles. Experimental hint to Higgs bosons are still missing.

2. The other approach is the string theory where the vibration modes of the string correspond to different elementary particles and where the vibration frequency multiplied by Planck's constant correspond to particle mass. Although the string theory is appealing, there are too many possibilities in it and there is still no one to one correspondence to the known phenomenology of particle spectrum.

As known also, there is still no quantum theory, describing gravitational interactions.

Let's assume that someone develops a new theory based on the concept of an underlying medium of whatever nature or structure where the known particles/fields correspond to "phonon-like"(quantized) excitations of this medium. And let's further assume that the theory can produce the known particle spectrum including corresponding masses (as the frequency $\omega$ at $k = 0$ in the dispersion relation $\omega(k)$ that fits well the known relativistic relationship $E(p)$ at least in the region for experimentally achievable energies) and types of excitations in the spectrum suggesting to represent gravitons. And let's assume further that the model successfully describes the known quantum processes including gravitational interaction as interactions between these "phonon-like" excitations.

What should we think of such a theory? Could we throw away such a theory only because it refers to a preferred frame? Obviously not, because in the new alternative theory there would be important additional properties explained successfully, that couldn't be explained previously. In this case, the domain of successfully covered phenomena would have changed in favor of the new theory (greater nominator "DP" in 6.1) so that it may be more preferable when compared to relativistic view as far as quantum processes are considered. However, even in this case, since Einstein's General Theory of Relativity would still be the successful theory at astronomic and cosmic scales, both viewpoints would continue to coexist. Only if the new theory in its limits at astronomic and cosmic scales can reproduce results that are at least as successful as the descriptions of General relativity, we may consider such a neo-Lorentzian view to replace relativistic view. As Villani and Arruda point out (1998 pp. 92) the development of general theory of relativity and its success had great influence in convincing the scientific community to adopt the relativistic picture and to abandon pre-relativistic or Lorentzian picture. Thus a discussion of the prospects for the compatibility between general theory of relativity and macroscopic limits of a quantum theory based upon neo-Lorentzian perspective is an interesting issue but it goes far beyond the main objective of the paper namely how to use Lorentzian route to Lorentz transformations as a didactical instrument to allow a smooth transition to relativistic picture as outlined in section 4 by addressing the didactical problems associated with the conventional introduction to special relativity listed in section 2.

## 7. Conclusions

There are several conceptual problems associated with conventional approach to introduce STR (Section 2). The historical route to Lorentz transformation may be a valuable instrument to resolve these pedagogical difficulties associated with conventional method listed in Section 2. To understand Einstein's viewpoint the student should be given the opportunity to realize that the concept of distant simultaneity is not something that nature provides us directly by observation and that the necessity of some type of convention is inevitable. Only by following this historical route up to its end, the student may fully appreciate the value of Einstein's approach which was one of the most profound applications of Occam's razor. However, we must be careful that Occam's razor should not cut too deeply or too permanently. We must keep namely in mind that it is not a particular application of Occam's razor at some point of history that should be considered as absolute truth, but it is the principle Occam's razor itself or better its generalized form (6.1 in section 6) that should be considered as the useful measure to decide between theories provided it is always reevaluated under new conditions. Thus, if some new theory may have an explanatory power beyond existing theories it may well be allowed to have some aspects that wouldn't appear to be in accordance with some previous application of Occam's razor.

Appendix-1

Conceptual schema

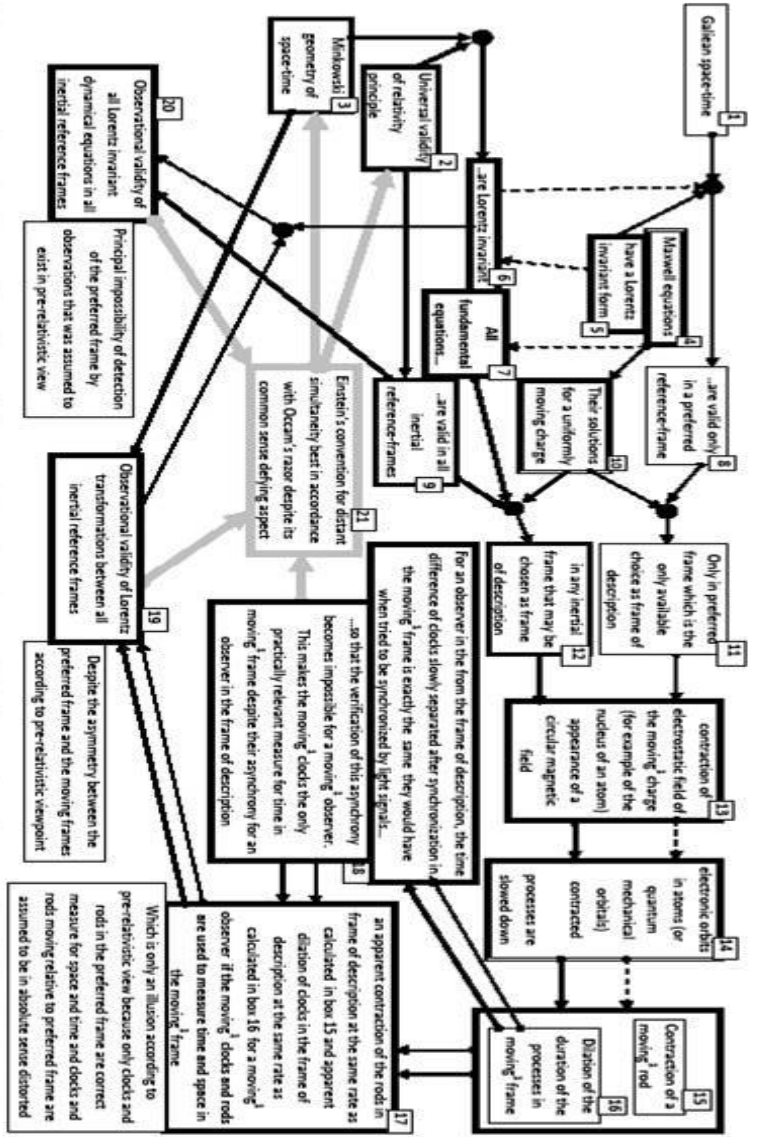

## Appendix-2. Alternative derivation of Lorentz transformations

Let's assume that S is the preferred frame where the Lorentz invariant fundamental equations are valid and we have one clock at the origin of reference-frame S and two clocks at the origin of reference frame S´ that is moving relative to S with velocity v. We assume that measuring rods in frame S´ are contracted by factor γ when compared to measuring rods in S because of Lorentz-Fitzgerald contraction and clocks in S´ run slower than the clocks in S by factor γ because of Larmor dilation. All three clocks are synchronized to t = 0 while the origin of S´ is passing by S. One of the clocks in S´ remains at the origin of S´ (clock-1) and the other one (clock-2) is separated from clock-1 with a separation velocity Δv (as measured in S) that is small compared to v so that its velocity relative to S is v+Δv. At a certain time when the reading of the clock at the origin of S is t then the reading of the clock-1 is $t/\gamma_{(v)}$ and the reading of clock-2 is $t/\gamma_{(v+\Delta v)}$. Since Δv is small when compared to v and since we want to consider the limit case where Δv goes to zero, we can take the first two terms in the series expansion of $t/\gamma_{(v+\Delta v)}$. For $\gamma_{(v)}$ we simply write γ.

1      $t/\gamma_{(v+\Delta v)} = t/\gamma + \Delta v \; d/dv \, (t/\gamma)$

We obtain

2      $t/\gamma_{(v+\Delta v)} = t/\gamma - t \, \Delta v \, \gamma \, v/c^2$

t Δv is the distance Δx between clock-1 and clock-2 (as measured with meter sticks in frame S).

3      $t/\gamma_{(v+\Delta v)} = t/\gamma - \Delta x \, \gamma \, v/c^2$

Thus the inevitable reading difference between clock-1 and clock-2 after slow separation in the limit case when the separation velocity goes to zero is $\Delta x \gamma v/c^2$ or $\Delta x´ v/c^2$ if Δx´ is the distance is measured with Lorentz-Fitzgerald contracted meter sticks in S´. This reading difference has exactly the same value one would obtain if one would attempt to synchronize them by light signals. In other words the reading difference of distant clocks that are slowly separated after synchronization in the moving frame has exactly such a value that we would obtain the same value c if we would try to measure the one way velocity of light in the moving frame in opposite directions with these asynchronous clocks. Please note that for v = 0 there is no contribution from first order derivative with respect to v in the series expansion and there are contributions only from terms with higher derivatives with respect to v. This means that clocks separated in the preferred frame (v = 0) according to the same procedure can become arbitrarily close to be synchronous if the separation velocity goes to 0. Since clock-1 is at the origin of S´, the coordinate of clock-2 in S´ is x´= Δx´. Thus the reading of a clock t´ in S´ at an arbitrary position x´ is given by:

4      $t´ = t/\gamma - x´v/c^2$

Thus if t is independent of x (which means that clocks distributed in S are synchronous) then t´ appears to be a linear function of x´. By eliminating x´ above using the Lorentz coordinate transformation Formula x´ = γ (x - vt), one obtains the Lorentz transformation formula $t´ = \gamma( t – xv/c^2)$. Since the Lorentz coordinate transformation Formula x´ = γ (x - vt) can be obtained at the very beginning using only the contraction hypothesis alone, complete Lorentz transformations can be obtained using the Lorentz Fitzgerald contraction and Larmor dilation without asserting the counterintuitive concept of invariance of light speed in advance in an introductory course. However according to the interpretation behind this derivation, only clocks distributed in S are synchronous in absolute sense (t does not depend on x) and clocks in the moving frame S´ are asynchronous in absolute sense (eq.4). If we derive the inverse formula we realize that we can obtain following linear function for t(x) similar to eq. 4 :

5      $t = t´/\gamma + xv/c^2$

Here t seems to depend linearly on x however this is illusory according to this interpretation because the linear dependence of t´ on position in the first term on the right side of eq. 4 has such a form that it compensates exactly the second term so that at the end t doesn't depend on x. Only after introducing the concept of relativity of simultaneity, namely by stating that for an observer in S´, clocks in S´ can be declared to be synchronous (namely by stating that for an observer in S´, t´ can be declared as being independent on x´ so that this time the eq.5 represents an actual linear dependence of t on x) the transformation formula obtain their relativistic interpretation.